\documentclass[epjCONF]{svjour}
\usepackage{graphics}
\usepackage[varg]{txfonts} 
\usepackage[latin1]{inputenc}
\session-title{1st International Conference on New Frontiers in Physics}
\begin{document}

\title{Limits on compact halo objects as dark matter from gravitational 
microlensing}

\author{Philippe Jetzer}

\institute{
Institute of Theoretical Physics\\
           University of Z\"urich, 
Winterthurerstarsse 190, CH-8057 Z\"urich, Switzerland\\
E-mail: jetzer@physik.uzh.ch}

\abstract{
Microlensing started with the seminal paper by Paczy\'nski
in 1986 \cite{pacz}, first with observations 
towards the Large Magellanic Cloud 
and the galactic bulge. Since then many other targets have been observed
and new applications have
been found. In particular, it turned out to be a powerful method to detect 
planets in our galaxy and even in the nearby M31.
Here, we will present some results obtained so far by microlensing
without being, however, exhaustive.}

\maketitle

\section{Introduction}

Since Paczy\'nski's original proposal \cite{pacz}
gravitational microlensing has been proven to be
a powerful tool for the detection of the dark matter
component in galactic haloes in form of MACHOs
and for determining the structure of
the Galaxy itself. Microlensing allows the detection of MACHOs
(Massive Astrophysical Compact Halo Objects) located in the Galactic
halo in the mass range  $10^{-7} < M/M_{\odot} <  1$
\cite{derujula,kn:Derujula1}, as well as MACHOs in the disk or bulge of our
Galaxy \cite{kn:Paczynski1991,kn:Griest2}. In September 1993, the
French collaboration EROS (Exp\'{e}rience de Recher\-che d'Objets Sombres)
\cite{kn:Aubourg} announced the discovery of two microlensing
candidates, and the American--Australian collaboration MACHO (for the
collaboration they use the same acronym as for the compact objects) of
one candidate \cite{kn:Alcock} by monitoring several millions of stars
in the Large Magellanic Cloud (LMC). 
The OGLE collaboration found one event towards the galactic bulge
\cite{udalski}. 
Since these first discoveries substantial
progress has been done: many more observations have been carried out
towards the LMC and the Small Magellanic Clouds (SMC), the galactic bulge,
the Andromeda galaxy (M31), some globular clusters. Microlesing
has develop into an important tool for the detection of extrasolar
planets in our Galaxy and even in M31. 
For an update on this topic we refer, e.g., to
the review by Dominik \cite{kn:dominik}.
It is also a useful tool to study stellar
atmospheres \cite{pol}. In the following we will present some of the results 
obtained so far towards different targets. This is, however, not an 
exhaustive review (see e.g. \cite{grg}), rather a 
personal choice of topics related to our own work in this field.

\section{Microlensing towards the Large Magellanic Cloud}

Since the first detection 
the MACHO team reported the observation of 13 to 17 events by
analyzing their $\sim$ 6 years of LMC data, leading to an
optical depth of $\tau = 1.0^{+0.3}_{-0.3}
{\times} 10^{-7}$ \cite{kn:bennett1}. Accordingly, this would imply a most probable
Galactic halo mass fraction of $\sim$ 20\% in the form of MACHOs with a most
likely mass in the range 0.15 - 0.9 $M_{\odot}$ depending on the
halo model. The MACHO value, however, is still substantially
higher then the upper limit of $\tau < 0.36 {\times} 10^{-7}$
presented by the EROS collaboration \cite{tisserand}, based on the
analysis of 33 million stars towards the LMC followed during $\sim$ 7
years. Recently, also the OGLE collaboration has published its
results based on almost 8 years of observations and
finds $\tau = (0.16 \pm 0.12) {\times}
10^{-7}$, concluding that at most 7\% of the galactic halo mass
could be due to MACHOs. However, most probably the events they
found are due to self-lensing or lenses located in the disk of our
own Galaxy, in which case the upper limit gets even more
stringent, ruling thus the MACHOs almost, if not completely out
\cite{kn:wyrzykowski1}, leaving it rather difficult to explain the results of
the MACHO collaboration \cite{kn:calchi+mancini}. Hopefully further observations
will help clarifying this issue. The OGLE collaboration published
recently its observations towards the 
SMC, where they found 3 good candidates yielding an optical
depth of $\tau = (1.30 \pm 1.01) {\times} 10^{-7}$, which could be
consistent with self-lensing in the SMC and a contribution from
our own Galactic disc \cite{kn:wyrzykowski}. However, a more
detailed analysis is needed.

\subsection{LMC self-lensing event rate}

The hypothesis for a self-lensing component
was discussed by several authors 
\cite{sahu,salati,evans98,zhao}.
The analysis of Jetzer et al. \cite{jetzer02} and 
Mancini et al. \cite{mancini} 
has shown that probably the observed
microlensing events, in particular the ones by the MACHO
collaboration, are distributed among different galactic components (disk,
spheroid, galactic halo, LMC halo and self-lensing). This means that
the lenses do not belong all to the same population and their
astrophysical features can differ deeply one another.

\noindent
Some of the events found by the MACHO team are most probably due
to self-lensing: the event MACHO-LMC-9 is a double lens with
caustic crossing \cite{alcock00b} and its proper motion is
very low, thus favouring an interpretation as a double lens within
the LMC. The source star for the event MACHO-LMC-14 is double \cite{alcock01b} 
and this has allowed to conclude that the
lens is most probably in the LMC. The expected LMC self-lensing
optical depth due to these two events has been estimated to lie
within the range \cite{alcock01b} $1.1-1.8\times10^{-8}$,
which is still below the expected optical depth for self-lensing
even when considering models giving low values it.
The event LMC-5 is due to a disk lens \cite{alcock01c} and
indeed the lens has even been observed with the HST. 
The other stars which have been
microlensed were also observed but no lens could be detected, thus
implying that the lens cannot be a disk star but has to be either
a true halo object or a faint star or brown dwarf in the LMC
itself.

\begin{figure}
\resizebox{\hsize}{!} {\includegraphics{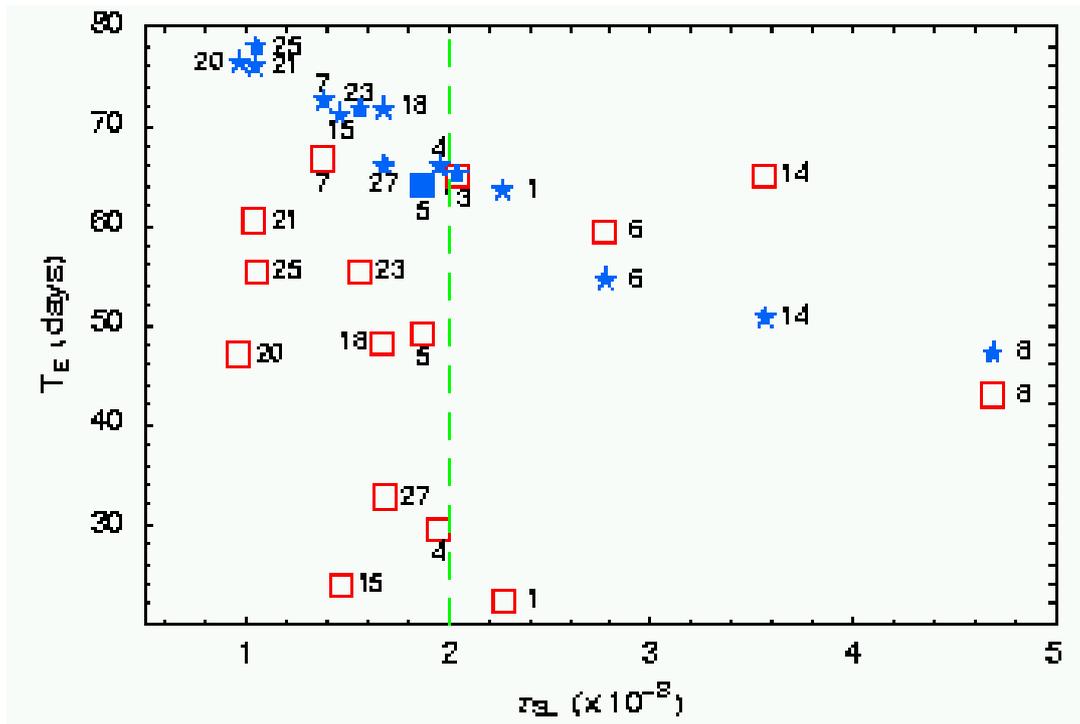}}
{\caption{Scatter plot of the observed (empty boxes) values of the
Einstein time and of the expected values of the median
$T_{\mathrm{E},50\, \%}$ (filled stars), with respect to the
self-lensing optical depth evaluated along the directions of the
events. From \cite{mancini}.}
\label{tevstau}}
\end{figure}

\noindent
We have calculated the self-lensing distributions
of the rate of microlensing events with respect to the Einstein
time $T_{\mathrm{E}}$, along the lines of sight towards 14
events found by the MACHO collaboration. 
With these distributions we have calculated 
the median $T_{\mathrm{E},\,50\,\%}$ values of the Einstein time.
In Fig. \ref{tevstau} we report on the $y$--axis the observed
values of $T_\mathrm{E}$ (empty boxes) as well as the expected
values for self-lensing of the median
$T_{\mathrm{E}\,,50\,\%}$ (filled stars) evaluated
along the directions of the events. On the $x$--axis we
report the value of the self-lensing optical depth $\tau_{\mathrm{SL}}$ calculated
towards the event position; the optical depth is  growing going
from the outer regions towards the center of LMC. An interesting feature
emerging clearly is the decreasing trend of the  expected
values of the median $T_{\mathrm{E}\,,50\,\%}$, going from the
outside fields with low values of $\tau_{\mathrm{SL}}$ towards the
central fields with higher values of $\tau_{\mathrm{SL}}$. The
variation of the stellar number density and the flaring of the LMC
disk certainly contributes to explain this behaviour.

\noindent
Based on these results we tentatively identify two subsets of events: the nine
falling outside the contour line $\tau_{\mathrm{SL}} = 2 \times
10^{-8}$  and the five falling inside. 
We note that, at glance, the two clusters have a clear-cut
different collective behaviour: the measured Einstein times of the
first $9$ points fluctuate around a median value of 48 days, very
far from the expected values of the median $T_\mathrm{E}$, ranging
from 66 days to 78 days, with an average value of 72 days. On the
contrary, for the last 5 points, the measured Einstein times
fluctuate around a median value of 59 days, very near to the
average value 56 days of the expected medians, ranging from 47
days to 65 days. Let us note, also, the somewhat peculiar position
of the event LMC--1, with a very low value of the observed
$T_{\mathrm{E}}$; 
most probably this event is homogeneous to the set at left of the
vertical line in Fig. \ref{tevstau} and it has to be included in
that cluster.
This plot suggests that the cluster of 
the $9$ events including LMC--1  belongs, very
probably, to a different population.
Clearly, given the large uncertainties and the few
events at disposal it is not yet possible to draw sharp conclusions,
nevertheless it is plausible that up to 
3-4 MACHO events might be due to lenses in LMC, which are most probably
low mass stars, but that hardly all events can be due to
self-lensing. The most
plausible solution is that the events observed so far by the MACHO team
are due to
lenses belonging to different intervening populations: low mass
stars in the LMC, in the thick disk, in the spheroid and some true
MACHOs in the halo of the Milky Way and the LMC itself.

\noindent
Moreover, it 
emerges that the spatial distribution of the events observed so far
shows a near--far asymmetry, which turns out to be compatible with the
optical depth calculated for LMC halo objects \cite{mancini}. The expected
characteristics of the lenses belonging to the MACHO population of the
Milky Way halo do not match the observed ones. This suggests that this
contribution can not explain all the observed candidates as well. 
Accordingly, we challenged the view that the MACHO dark halo 
fraction of both the
Milky Way and the LMC halos are equal, and indeed we showed that for a
MACHO mass in the range 0.1-0.3 M$_\odot$ the LMC halo fraction can be
significantly larger than the Milky Way one \cite{calchi1}. In this
perspective, it might well be that about half of the observed events
are due to lenses located in the LMC halo.

\section{Microlensing towards the Galactic bulge region}

\noindent To date, the MACHO \cite{kn:MACHO}, EROS \cite{kn:afo},
MOA and OGLE \cite{kn:ogle} collaborations found several thousands
microlensing events towards the Galactic bulge. Considering as
sources only the clump giant stars the MACHO team finds a value
for the optical depth $\tau = 2.17^{+0.47}_{-0.38}{\times}
10^{-6}$ \cite{kn:pop}. Similar values for the optical depth have
been found by the OGLE \cite{sumi} and the EROS \cite{hadamache}
teams \cite{kn:moniez}. The MACHO and EROS collaborations found also several events
towards the spiral arm regions \cite{kn:eros}. These results are
important for studying the structure of our Galaxy and the stellar
mass function \cite{kn:jetzer1}. Recently, the MOA collaboration
\cite{kn:sumi1} found a population of unbound or distant planetary
mass population (with masses of a few Jupiter mass) and could
better determine the slope of the stellar mass function at low
masses.

\noindent
We addressed the problem of the shape of  
the initial mass function of the Galactic bulge using the microlensing
observations towards the Galactic center by the MACHO, 
EROS and OGLE collaborations \cite{kn:grenacher,kn:jetzer1}.   
In particular, we considered the duration distribution of the 
microlensing events. Assuming a power law for the initial mass function
we studied the slope, as given by the power law index, both in the brown dwarf
and in the main sequence ranges. Moreover, we compared the 
observed and theoretical, based on bulge modelling,
optical depth profiles, which turn out to be in quite good
agreement \cite{kn:jetzer1}.

\noindent
Furthermore, using the current microlensing and dynamical observations of the Galaxy one can set
interesting constraints on the dark matter local density and profile slope towards the galactic centre.
Assuming state-of-the-art models for the distribution of baryons in the Galaxy, we found, for instance, that the most
commonly discussed dark matter profiles (Navarro-Frenk-White and Einasto) are consistent
with microlensing and dynamical observations, while extreme adiabatically compressed profiles are
robustly ruled out. When a baryonic model that also includes a description of the gas is adopted,
our analysis provides a determination of the local dark matter density, $\rho_0 = 0.20 - 0.56~ GeV/cm^3$
at 1$\sigma$, that is well compatible with estimates given in the literature based on different techniques \cite{bertone}.
Clearly, once more microlensing data will be gathered and thus better galactic models will
be constructed the above estimates might become more precise and thus also more interesting.

\begin{figure}[ht]
\resizebox{\hsize}{!} {\includegraphics{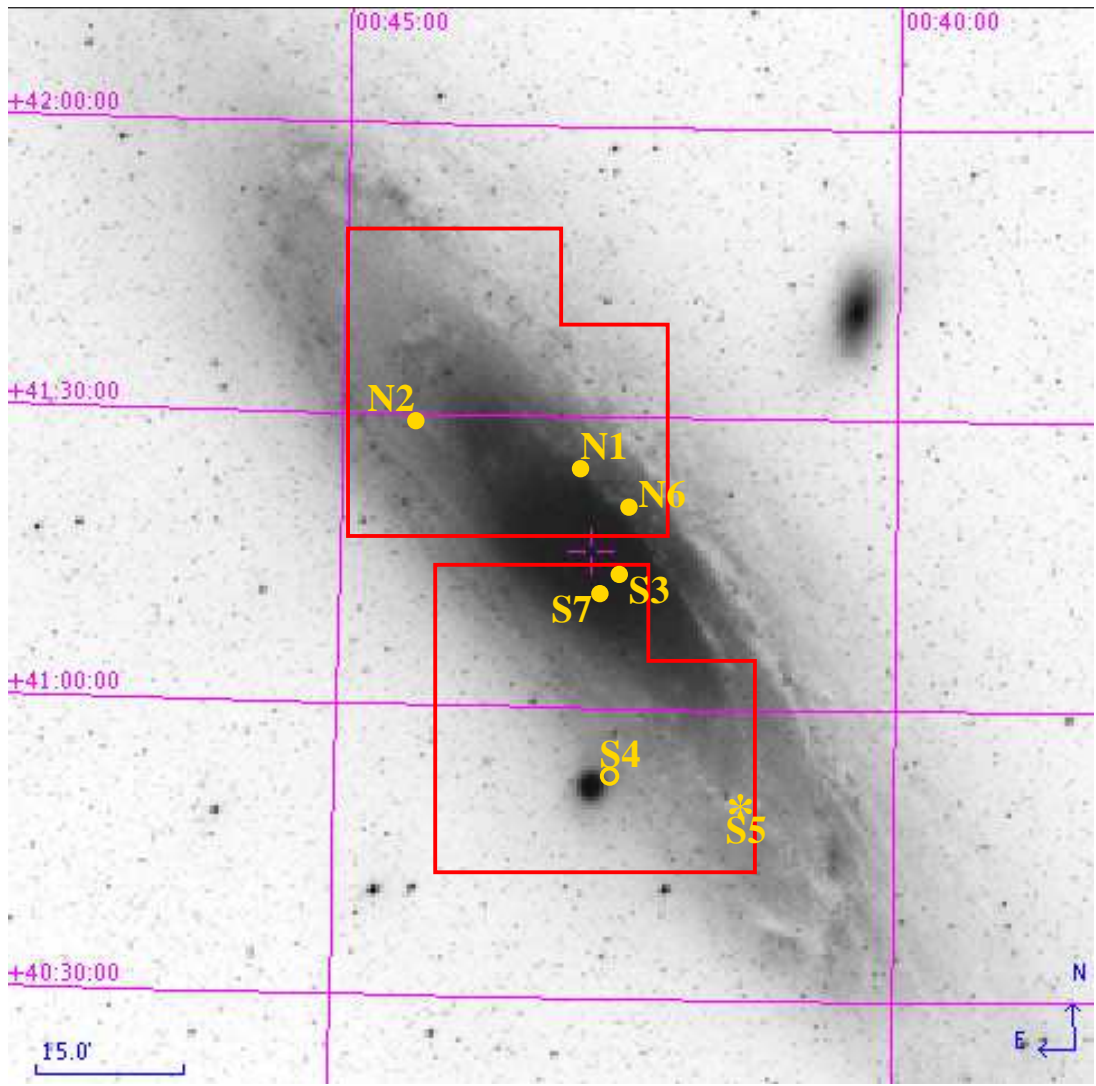}}

\caption{Positions of some 
microlensing events found so far projected on M31. The solid lines
show the contour of the observed fields. Note that the event
denoted by S4 almost on the line of sight of M32, a satellite
galaxy of M31. From \cite{calchi}.}
\label{int-pos}
\end{figure}

\begin{figure}[ht]
\resizebox{\hsize}{!} {\includegraphics{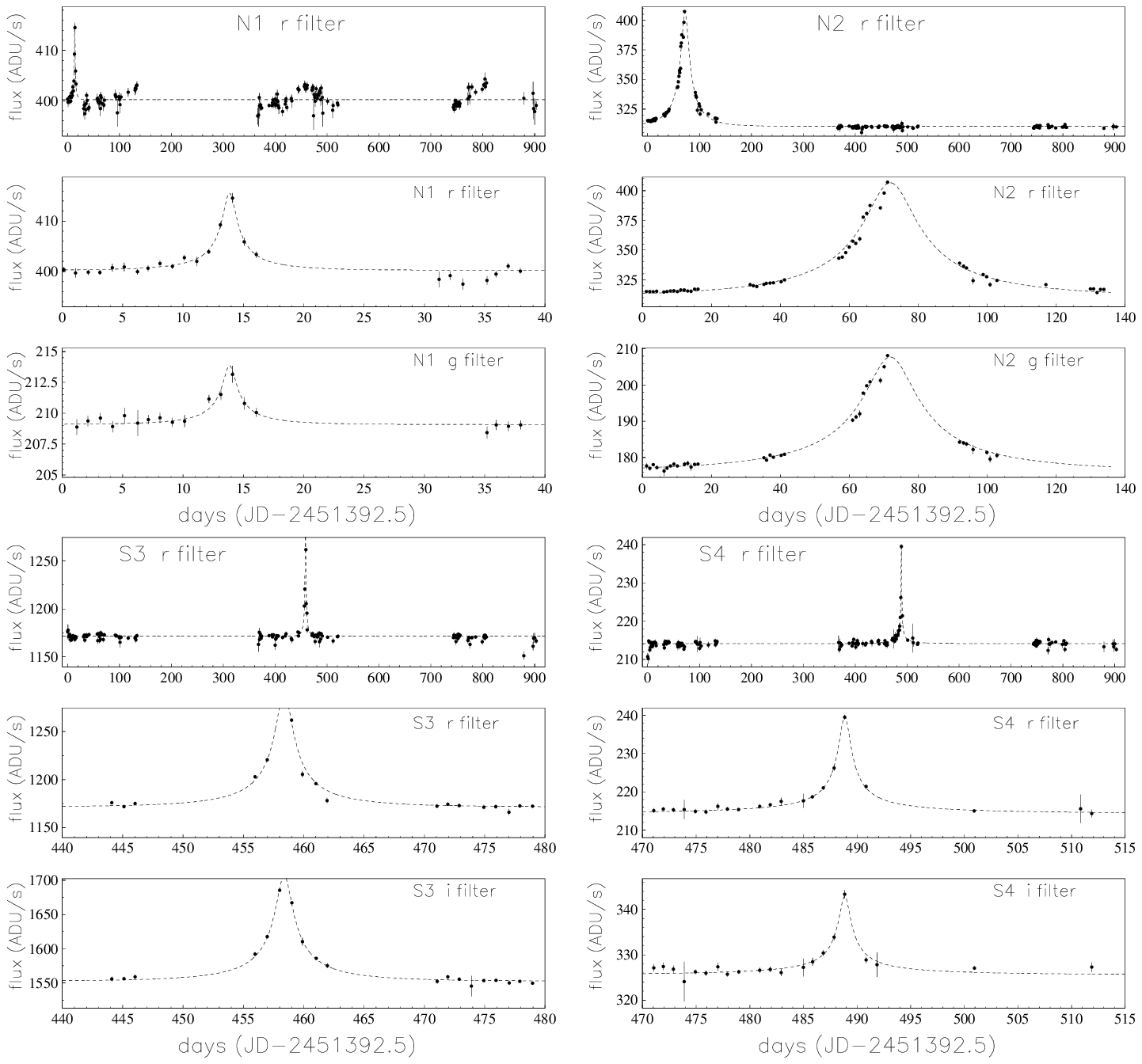}}
\caption{Three years data light curves for 4 microlensing events
found by the POINT-AGAPE collaboration. From \cite{calchi}.}
\label{int4}
\end{figure}

\begin{figure}[ht]
\resizebox{\hsize}{!} {\includegraphics{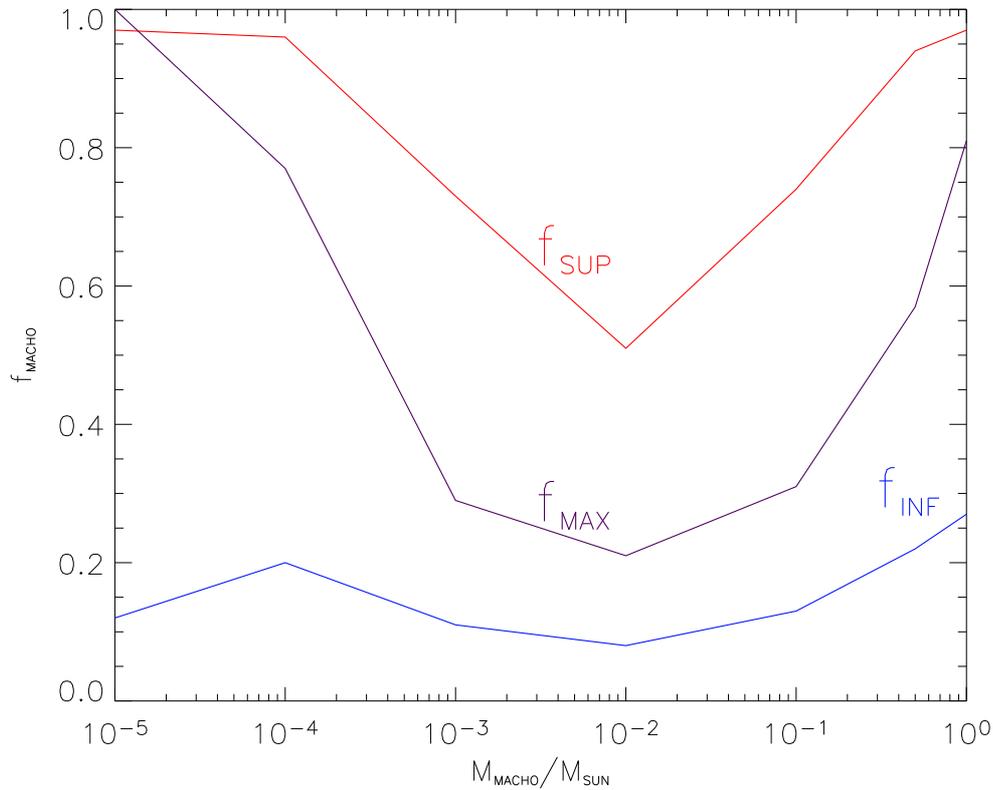}}
\caption{Most probable value, upper and lower 95\% CL limit for the halo 
fraction as a function of the MACHO mass in solar mass units, based on the POINT-AGAPE
results. From \cite{calchi}.}
\label{int4}
\end{figure}

\begin{figure}[ht]
\resizebox{\hsize}{!} {\includegraphics{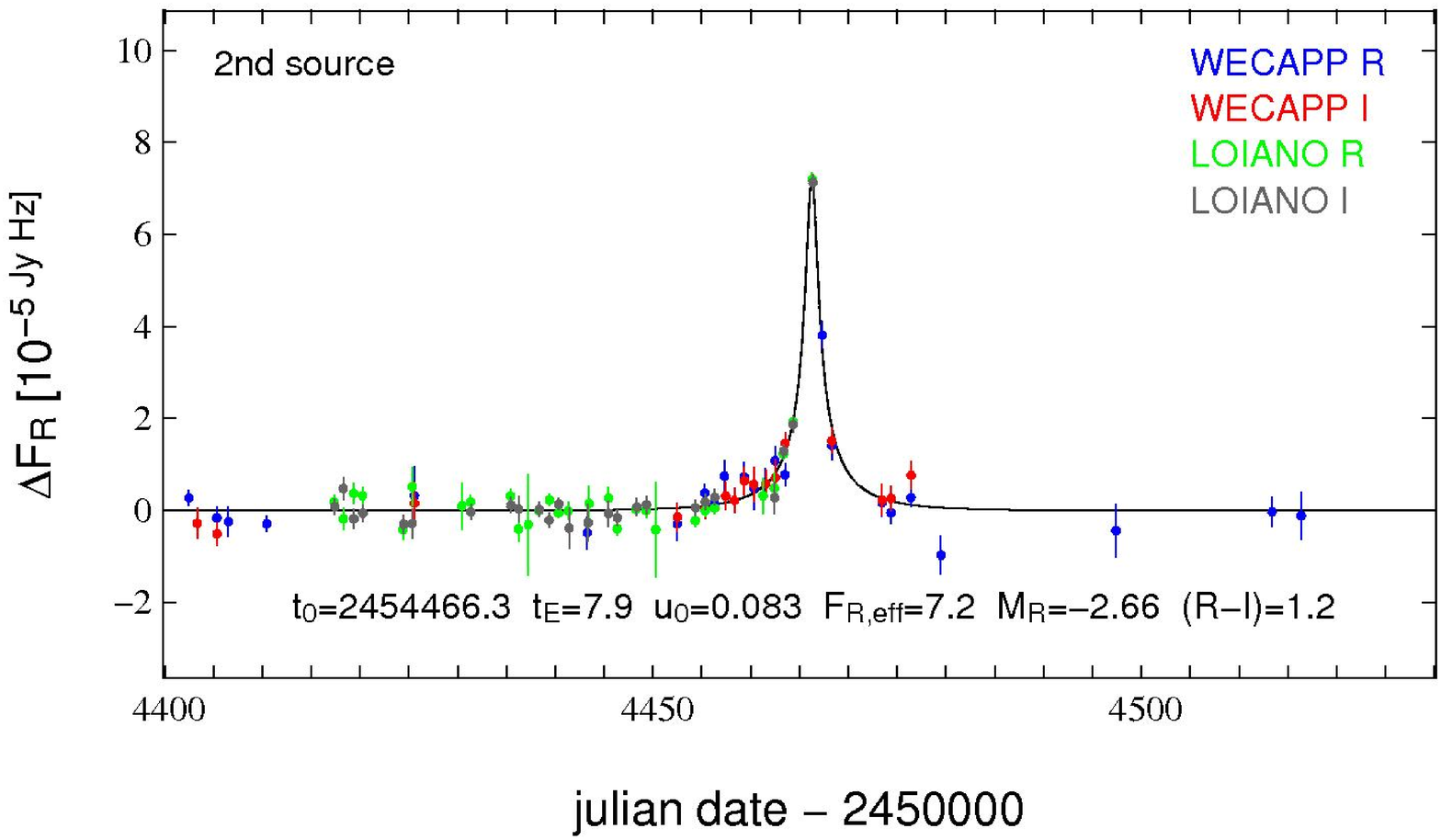}}
\caption{OAB-N2 light curve as obtained by combining PLAN and WeCAPP data. From
\cite{jetzer1}.}
\label{int4}
\end{figure}

\section{Microlensing towards globular clusters}

\noindent
In the last years we analysed the large set of microlensing events detected
so far towards the Galactic center with the purpose of investigating whether some 
of the dark lenses are located in Galactic globular clusters \cite{straessle}.
We found that in some cases the events might indeed be due to lenses
located in the globular clusters themselves.
We got a rough estimate for the average lens mass
of the subset of events being highly aligned with Galactic globular cluster centers
and found that, under reasonable assumptions, 
the deflectors might most probably be either brown dwarfs, M-stars or  
stellar remnants \cite{deluca}.

\noindent
In 2000 July/August a microlensing event occurred at a distance of 2.33 arc minutes
from the center of the globular cluster M22 (NGC 6656), observed against the
dense stellar field of the Milky Way bulge. 
In order to check the hypothesis that the lens belongs to the globular
cluster we made a dedicated observation,
using the adaptive optics
system NACO at the ESO Very Large Telescope to resolve the two objects - the lens and the source - that
participated in the event. The position of the objects
measured in July 2011 was in agreement with the observed relative proper motion
of M22 with respect to the background bulge stars. Based on the brightness of
the microlens components we found that the source is a solar-type star located at
a distance of 6.0 $\pm$ 1.5 kpc in the bulge, while the lens is a
0.18 $\pm$ 0.01 $M_{\odot}$ dwarf
member of the globular cluster M22 located at the 
known distance of 3.2 $\pm$ 0.2 kpc from the Sun \cite{pawel}.
This is indeed the first confirmed microlens in a globular cluster. 
It would be desirable to get more observation of globular clusters,
indeed  few events 
could already help very much to constrain the low mass star content 
and thus to get a clear mass budget and information on the stellar mass function of the globular cluster.

\section{Microlensing events towards M31}

Microlensing searches towards the Andromeda galaxy (M31) have also
been proposed \cite{kn:Crotts,kn:Baillon,kn:Jetzer}. In this case
one has to use the so-called ``pixel-lensing'' method, since
the source stars are in general no longer resolvable. This makes the
subsequent analysis more difficult, however, it allows M31 and other
objects, such as M87 \cite{kn:sbaltz}, to be used as targets
\cite{kn:calchi}. For
information on the shape of the dark halo, which is presently unknown,
it is important to observe microlensing in different directions
as can be done towards M31.
Several teams performed searches, in particular the French AGAPE \cite{kn:Z1},
the POINT-AGAPE \cite{calchi}, the MEGA \cite{kn:mega1}, the WeCapp
\cite{kn:wecapp,wecapp1} and the PLAN (Pixel Lensing ANdromeda) \cite{plan1,plan}. All
groups found some candidate events consistent with
microlensing.
As a main result, a high-threshold analysis of the 3 years of POINT-AGAPE data yielded
6 bright, short--duration microlensing events.
The observed events are more
 than expected from self lensing alone,
thus leading to the conclusion that at least 20\% of the  halo
mass in the direction of M31  should be in the form of MACHOs, if
their average mass lies in the range   0.5-1 M$_\odot$
\cite{calchi}. Also the MEGA group presented their results with
the detection of 14 candidate events \cite{kn:mega1}. The observed
timescale distribution of the events seems to be consistent with
halo lensing dominating in the outer parts of M31. 

\noindent
Within the PLAN collaboration we found during the 2007
season  two candidate microlensing events \cite{plan}. 
One event (OAB-N2) was at the end
of the observing period, but fortunately the WeCAPP collaboration made observations afterwards.
We thus shared the data and the combined analysis shows that with the better coverage the
event is nicely confirmed.
The improved photometry enabled a detailed analysis
in the event parameter space including the effects due to finite source size. The combined results of
these analyses allowed us to put a strong lower limit on the lens proper motion. This outcome favours
the MACHO lensing hypothesis over self-lensing for this individual event and points the way towards
distinguishing between the MACHO and self-lensing hypotheses from larger data sets \cite{jetzer1}.

\noindent
The PAndromeda team (a following up of the WeCapp collaboration)
using the Pan-STARRS survey done with a 1.8 meter telescope
found recently 6 microlensing events towards M31 \cite{rif}.
The interpretation of these results in terms of limits on the dark
matter in form of MACHOs in M31 has, however, not yet been done. 
More observations are needed to clarify these preliminary
results, nonetheless today it is established that the pixel-lensing 
method works.

\subsection{Pixel-lensing as a way to detect extrasolar planets in M31}

\noindent
We studied also the possibility to detect extrasolar planets in M31 through pixel-lensing observations,
similar to what is done by classical microlensing techniques towards the galactic center \cite{ingrossop}. 
Using a Monte Carlo approach, we selected the physical parameters of the binary lens system, a star 
hosting a planet, and we calculated the pixel-lensing light curve taking into account the finite source effects. 
Indeed, their inclusion is crucial since the sources in M31 microlensing events are mainly giant stars. 
Light curves with detectable planetary features are selected by looking for significant deviations from the 
corresponding Paczy\'{n}ski shapes. We found that the time scale of planetary deviations in light curves increase 
(up to 3-4 days) as the source size increases. This means that only few exposures per day, depending also on the 
required accuracy, may be sufficient to reveal in the light curve a planetary companion. Although the mean planet 
mass for the selected events is about $2 M_{\rm {Jupiter}}$, even small mass planets ($M_{\rm P} < 20 M_{\oplus}$) 
can cause significant deviations, at least in the observations with large telescopes. However, even in the former case, 
the probability to find detectable planetary features in pixel-lensing light curves is at most a few percent of the 
detectable events, and therefore many events have to be collected in order to detect an extrasolar planet in M31. 
Our analysis also supports the claim that the anomaly found in the candidate event PA-99-N2 towards M31 
\cite{point03a} can be explained 
by a companion planetary object of about 6 Jupiter masses orbiting the lens star \cite{ingrossop}.

\section{Conclusions} 

Microlensing has proven in these years to be a very powerful
method in various areas of astrophysics: from the determination of the dark
matter content in form of MACHOs in the halo of our Galaxy
and in the one of M31, for the study of the structure of our Galaxy,
the determination of the stellar mass function, particularly
in its low mass end, to the detection of planets in the Galaxy and
even in the nearby M31.
Progress has also been done in the study of stellar atmospheres
or the low mass star content of globular clusters.
All these topics and maybe others will greatly benefit from
more observations and data which will surely come in the next years.
Discussion are also underway to use satellites like Euclid  
for performing microlensing observations towards the
Galactic center with the aim to find planets.
Microlensing is now a mature technique with many applications which will help
solving current astrophysical problems.

------------------------------------------------------------------

\end{document}